\documentclass[12pt,a4paper]{iopart}
\usepackage{color}
\usepackage{amsfonts}
\usepackage[ansinew]{inputenc}
\usepackage{epsfig}
\usepackage[pdftex]{hyperref}
\usepackage{enumitem}
\usepackage{setspace}
\usepackage{multirow}

\newcommand{\remark}[1]{}
\newcommand{\ketbra}[1]{\ensuremath{| #1 \rangle \langle #1 |}}
\newcommand{\ket}[1]{\ensuremath{|#1\rangle}}

\newcommand{\braket}[2]{\ensuremath{\langle #1|#2\rangle}}
\newcommand{\Ca}{\ensuremath{^{40}\textrm{Ca}^+}\,}
\newcommand{\Sr}{\ensuremath{^{88}\textrm{Sr}^+}\,}
\newcommand{\Ba}{\ensuremath{^{138}\textrm{Ba}^+}\,}
\newcommand{\wavenum}{\ensuremath{\textrm{cm}^{-1}\,}}

\newcommand{\mum}{\ensuremath{\mu \textrm{m}\,}}

\fussy 
\begin{document}

\title{Ultrafast infrared spectroscopy with single molecular ions}
\author{Philipp Schindler}
\address{Institut f\"ur Experimentalphysik, Universit\"at Innsbruck, A-6020 Innsbruck, Austria}
\ead{philipp.schindler@uibk.ac.at}

\begin{abstract}

  We propose a method to investigate the vibrational dynamics of
  single polyatomic molecular ions confined in a Paul trap. Quantum
  logic techniques are employed to detect the recoil of single photon
  absorption events in the molecule via a co-trapped atomic ion. In
  particular, the recoil is mapped onto the electronic state of the
  atom which can be read out with high fidelity. This recoil detection
  serves as the basis for a pump-probe scheme to investigate ultrafast
  molecular dynamics, such as intra-molecular vibrational
  redistribution. The total recoil from the interaction with a
  sequence of ultrafast laser pulses with the molecular vibration is
  measured. This work discusses the experimental requirements and
  expected performance for multiple molecular ions with masses ranging
  from 17 to 165 Dalton.
\end{abstract}

\section{Introduction}

While the quantum state of individual atoms can be manipulated with
impressive
accuracy~\cite{Schindler2013,Wineland1995Experimental,Grimm1999},
control over single molecules has not reached a similar
level~\cite{Lemeshko2013}. The complex internal structure of molecules
is rarely suitable for the state preparation and measurement methods
that have been developed for atomic systems. In particular, direct
laser cooling of molecules has proven to be difficult and seems only
feasible in a few select molecular
species~\cite{Chandler2009,Shuman2010,Lemeshko2013}. Quantum logic
methods have been developed to improve control over ionized atomic and
molecular species that are inaccessible with standard optical-pumping
and laser-cooling methods~\cite{Schmidt2005}. This approach has proven
to be useful to enable atomic clocks~\cite{Schmidt2005} and also to
investigate diatomic molecular
ions~\cite{Chou2016,Wolf2016,Leibfried2012,Ding2012,Mur-Petit2012}. Here,
we discuss an extension of these methods onto the vibrational degree
of freedom of polyatomic molecules.

Intramolecular dynamics are usually investigated using time-domain
techniques in the picosecond to femtosecond
regime~\cite{Brinks2014}. In a typical experiment, a sequence of
ultrafast laser pulses with varying time delay is applied onto the
molecule. The available methods range from relatively simple
pump-probe procedures to complex multi-pulse
techniques~\cite{Brinks2014,Lemeshko2013}.
Such ultrafast experiments are performed routinely to measure the
absorption properties of neutral molecules in the gas phase.
Transferring these experiments directly onto trapped molecular ions is
difficult, because the Coulomb repulsion between the charged particles
limits the achievable density which leads to a poor signal to noise
ratio~\cite{Baer2010}. The signal to noise ratio can be improved by
either increasing the molecule density, or by increasing the interaction
strength between the molecule and the light field. For example,
velocity modulation spectroscopy in a dense plasma enables direct
absorption spectroscopy of charged particles but does not
allow mass selection of the ions and requires bright ion
sources~\cite{And2005,Polak1989}. Alternatively, cavity-enhanced
techniques increase the effective absorption cross-section, but are
incompatible with single ultrashort laser
pulses~\cite{Tan2006,Coe1989}. As a consequence, most available direct
spectroscopic techniques for molecular ions are destructive and
require disposing of the molecules after each experimental
cycle~\cite{Willitsch2017}.  The experiment in
reference~\cite{Kahra2012} is an example of a destructive detection
method combined with time-domain spectroscopy. There, electronic
transitions of diatomic molecular ions have been investigated with a
pump-probe technique that dissociates the molecule and detects the
fluorescence of the remaining atom.

Here, we describe a non-destructive single photon absorption detection
method by combining quantum-logic techniques with ultrafast optical
control: Each photon absorption event comes with a momentum kick that
alters the combined motional state of both co-trapped
particles~\cite{Hempel2013,Wan2014,Clark2010}. The change in momentum
is then mapped onto the electronic state of the atomic logic ion which
can be detected with high fidelity. Unfortunately, absorbing a single
photon results in a small momentum kick that cannot be detected if the
motional state of the ion crystal is close to its ground state. It has
been shown that single-photon sensitivity can be reached by using a
tailored, non-classical, motional state of the trapped
particles~\cite{Hempel2013}.  Here, we adapt this technique to detect
a single-photon absorption event on an infrared vibrational transition
in a polyatomic molecule.

For time-domain techniques, the coherent addition of recoils from
multiple laser pulses can be exploited, enabling measurements at a
much faster timescale than the recoil readout on the atom.
The simplest multi-pulse experiments are pump-probe
experiments, where an initial pump pulse initializes a vibrational
mode in a given state and the subsequent probe pulse is applied with
varying waiting time. The absorption of the probe pulse gives
information about the decay of the induced population from the
vibrational mode of interest. Such pump-probe experiments can serve as the
foundation for more complex multi-pulse time-domain experiments, such
as multi-dimensional spectroscopy~\cite{Brinks2014,Lemeshko2013}.

\begin{figure}[t]
\begin{center}
\includegraphics[width=.7 \textwidth]{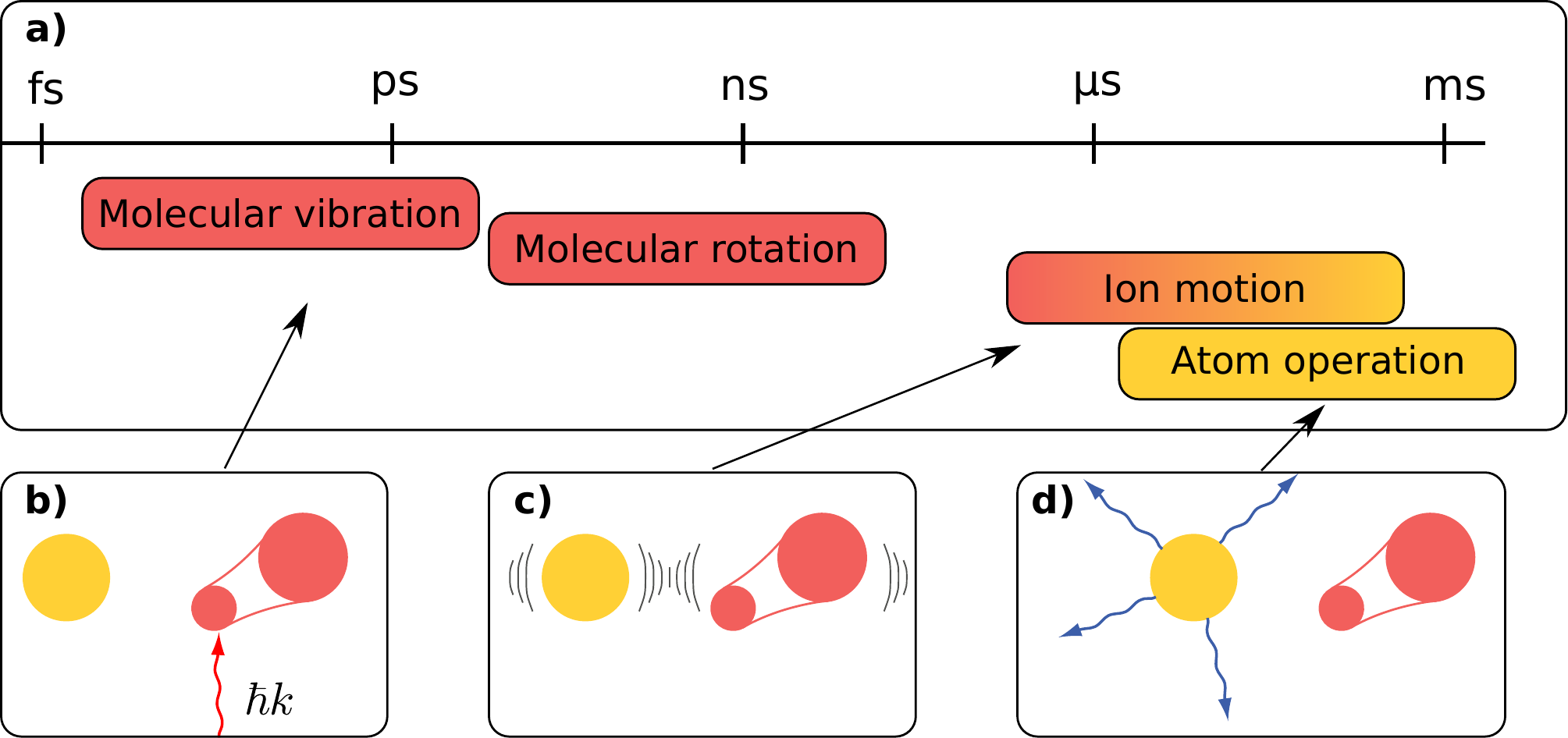}
\caption{\label{fig:abs_schem} a) Relevant timescales for the
  molecular quantum logic spectroscopy. The techniques are able to
  detect the recoil of the absorption of a single photon on the
  molecule (red) on the co-trapped atomic ion (yellow). The protocol
  can be sketched as follows: b) The molecule absorbs a single photon
  from a femtosecond laser pulse. c) The recoil of the absorbed photon
  alters the state of external motion of atom and molecule. d) The
  change in motional state is read-out via the electronic state of the
  atom. }
\end{center}
\end{figure}

The separation of timescales of molecular dynamics and atomic control
is crucial for the method (as sketched in figure~\ref{fig:abs_schem})
and can be summarized as follows:
\begin{itemize}
\item The molecule is irradiated with a single or multiple laser
  pulses at the femtosecond timescale. The absorption of photons leads
  to energy transfer from the light field to the molecule where each
  absorption event leads to an instantaneous change of the momentum of
  the molecule.
\item The Coulomb interaction transfers the momentum kick onto the
  co-trapped atomic ion. The relevant timescale for the momentum
  transfer throughout the ion crystal is given by the 
  ion-trap frequency of both ions which is in the microsecond regime.
\item The motional state of the atomic ion is mapped onto its
  electronic state which can be read out with high fidelity at the
  millisecond timescale.
\end{itemize}

This manuscript is structured as follows: The central method for
single-molecule absorption detection is covered in detail in
section~\ref{sec:absorption_spec}.
In
section~\ref{sec:pump_probe}, a pump-probe scheme to investigate
ultrafast intra-molecular dynamics using quantum logic techniques is
introduced. The requirements for an experimental realization are
discussed in section~\ref{sec:exper-requ} and the expected performance
is analyzed in sections~\ref{sec:expect-exper-perf}
and~\ref{sec:expect-exper-pump-probe}.

\section{Spectroscopic methods}

\subsection{Single photon absorption detection}
\label{sec:absorption_spec}

In the following, we introduce a technique to detect a single-photon
absorption event on a molecule via the co-trapped atomic logic ion. We
follow a semi-classical description: The light field of the
interrogation laser is treated classically whereas the internal
degrees of freedom and the motion of the particles are treated quantum
mechanically. We first use a simple model of the molecule that
includes only the ground state and the first excited state of a single
vibrational mode. The energy eigenstates of this two-level system are the
ground state $\ket{g}_m$ and the excited state $\ket{e}_m$.  The most
general pure state is then given by
$\ket{\Psi_{mol}} = c_e \ket{e}_m + c_g \ket{g}_m $. We will extend
this model to multiple vibrational modes with multiple levels below.
Analogously, the electronic state of the co-trapped atom is described
by a two-level system with basis states $\ket{g}_a$ and $\ket{e}_a$.
The motion of the two co-trapped particles in a linear ion-crystal is
described by a single-mode harmonic oscillator where a convenient
basis is spanned by the coherent states $\alpha_j$. Any pure motional
state can then be described as
$\ket{\Psi_{motion}} = \sum_j c_j \ket{\alpha_j}$.

We express the combined vibrational, atomic, and motional state of the
system as
$\ket{\Psi_{tot}}=\ket{\Psi_{mol}} \otimes \ket{\Psi_{atom}} \otimes
\ket{\Psi_{motion}}$.  The dynamics of the combined system can be
described as two two-level systems and a single harmonic oscillator,
yielding the Hamiltonian
\begin{equation}
  H_{mol} = \frac{\hbar}{ 2} (\omega_t a^\dagger a + \omega_v \sigma^{(m)}_z + \omega_a \sigma_z^{(a)})
  \label{eq:MolHamil2Level}
\end{equation}
with the motional trap frequency $\omega_t$, the transition frequency
of the vibrational mode~$\omega_v$, and the atomic transition
frequency $\omega_a$. The Pauli operators acting on the atomic or the
molecular system are denoted with the respective indices $(m)$ and
$(a)$.

The interaction of the  molecule and the vibrational mode with an oscillating light
field with frequency $\omega_l$ is governed by the Hamiltonian
\begin{equation}H_{el} =  \frac{\hbar \Omega}{ 2} (\sigma^{(m)}_+ e^{i \eta (a+a^\dagger)} e^{-i \omega_l t} + \sigma^{(m)}_-e^{-i \eta (a+a^\dagger)} e^{i \omega_l t} )
\label{eq:LightAtomHamil}
\end{equation}
where the position $\vec{x}$ of the molecule in the trap relative to the
electric field wavelength is expressed by the raising and lowering
operators $\vec{k} \cdot \vec{x}=\eta (a+a^\dagger)$. This introduces
the dimensionless Lamb-Dicke parameter which is defined as the product
of the light field's wave vector $\vec{k}$ and the ground state
wavepacket size of the harmonic oscillator $\vec{x}_0$:
\begin{equation}
  \eta = \vec{k} \cdot \vec{x}_0 = k \sqrt{\frac{\hbar}{2 m \omega_t}} \cos(\theta)
  \label{eq:LDDef}
\end{equation}
with $\theta$ being the overlap angle between the direction of the
oscillator's motion and the light's wave vector $k$, and $m$ the total mass
of the ion crystal.

The effect of the momentum kick, due to the absorption of a photon on
any of the trapped particles, can be described by applying a
displacement operator onto the motional mode~\cite{Hempel2013}. The
magnitude of the displacement is given by the Lamb-Dicke parameter
$\eta$. The timing of the absorption event in relation to the
oscillatory motion of the particles determines the phase of the
displacement operator.  Absorbing a photon on the molecule in its
vibrational ground state is thus described by the operator
\begin{equation}
  U_{abs} = D(\exp(i \phi_{sc})\eta) \sigma^{(m)}_+
  \label{eq:dis_operator}
\end{equation}
with $D$ being the displacement operator acting on the combined
motional state of the Coulomb crystal. Here, $\phi_{sc}$ describes the
scatter phase of the recoil event relative to the harmonic
oscillator's motion~\cite{Hempel2013}. A laser pulse that has fixed
timing with respect of the phase of the oscillatory motion and is
short compared to the oscillation frequency yields a fixed phase
$\phi_{sc} = \omega_t \tau$ with $\tau$ being the time difference
between a zero crossing of the harmonic oscillator and the laser
pulse. %

If we assume that the crystal is initially in its motional ground
state $\ket{\Psi_{motion}} = \ket{0}$ then the probability to detect
the absorption event by adding a single phonon is approximated by
$\eta^2$~\cite{Wineland1995Experimental}.
For transitions in the visible spectrum (around 15000\wavenum) and an
ion crystal with mass of around 80 Dalton, the Lamb Dicke factor is
below 10\% and thus the detection probability in the visible domain
is below 1\%. For infrared vibrational transitions with frequency of
3000\wavenum or below, the detection probability is even smaller. This
renders the faithful detection of a single absorption event impossible
when the initial motional state is the ground state.

It is therefore necessary to increase the detection sensitivity to
reach appreciable single-photon detection probability. This can be
accomplished by using a motional state of the ion crystal that is
entangled with the electronic state of the logic ion and tailored to
detect a small displacement.  Single-photon sensitivity has been
demonstrated in an experiment using an entangled Schr\"odinger cat
state of the motional mode of two distinct atomic
species~\cite{Hempel2013}. The particular state of the atom
and the common motion is
\begin{equation} \ket{\Psi_{cat}} = \frac{1}{\sqrt{2}} \biggl{(}\ket{+}_x \otimes \ket{\alpha} + \ket{-}_x \otimes \ket{-\alpha}\biggr{)}
\end{equation}\label{eq:cat_state}
where the electronic state of the atom is represented in the basis of
the $\pm1$ eigenstates of the $\sigma^{(a)}_x$ operator $\ket{\pm}_x$.

The creation procedure for this state is described and analyzed for
realistic experimental parameters in
section~\ref{sec:expect-exper-perf}. In the following, we assume
perfect state generation to introduce the concept of recoil
spectroscopy in molecular systems. The state of the combined system,
with the molecular vibration in its ground state, is then
$\ket{\Psi_{tot}} = \ket{\Psi_{cat}} \otimes \ket{g}_m$.

An absorption event of a single photon on the molecular ion displaces
the combined motional state of the atom and the molecule by the photon
recoil but does not alter the atom's electronic state. The action of
the photon recoil on the motional mode is given in
equation~(\ref{eq:dis_operator}). The state of the combined system
after an absorption event is then
\begin{equation} \ket{\Psi_{recoil}} = \frac{1}{\sqrt{2}} \biggl{(}\ket{+}_x \otimes \ket{\alpha + e^{i\phi_{sc}}\eta} + \ket{-}_x \otimes \ket{-\alpha + e^{i\phi_{sc}} \eta}\biggr{)} \otimes \ket{e}_m \; .
\end{equation}

The photon recoil is then mapped onto the electronic state of the atom
by applying the inverse of the creation
operation~\cite{Hempel2013}. After the inverse operation, the
displacement from the photon recoil causes a geometric phase between
the electronic states $\ket{\pm}_x$ of
\begin{equation}\theta_g = 2 \alpha \eta \sin \phi_{sc} \; .
\end{equation}
In analogy to light-force based quantum gates, this geometric phase
can be mapped to a bit-flip on the electronic state of the atom in a
Ramsey-type experiment~\cite{molmer}. The maximum sensitivity of the
protocol is reached if the timing of the interrogation laser pulse is
such that $\phi_{sc}=\pi/2$. In this case, the displacement acts along
the sensitive axis of the state $\ket{\Psi_{cat}}$. The probability to
detect a single photon absorption on the molecule is then
\begin{equation} p_{det} = \sin(2\alpha \eta)^2 \; . \label{eq:det_eff}\end{equation}

The presented absorption detection method is, in principle, applicable
without restriction to the investigated transition type if the
timescale of the light-molecule interaction is short compared to the
dynamics of the motional mode. However, the molecular vibration needs
to be described by multiple anharmonic oscillators instead of a simple
two-level system.  Such anharmonic oscillators are inconvenient for
numeric and analytic treatment and thus we consider two limiting cases
where the analysis can be simplified: (i)~The anharmonic frequency
shift is much larger than the spectral width of the applied laser
pulses. In this case, the mode can be approximated by a two-level
system, described by the Hamiltonian of
equation~(\ref{eq:MolHamil2Level}). (ii)~The spectral width of the
laser pulses is much larger than the anharmonic frequency shift,
yielding a harmonic oscillator.

We consider now multiple vibrational modes of the molecule in the
limit of small anharmonic frequency shifts. Thus, we express the state
of the molecule as a tensor product of all vibrational modes
$\ket{\Psi_{mol}} = \otimes_j \ket{ \nu_j}$ where the state of the
mode $j$ is described by $\ket{\nu_j} = \sum_k c_{k,j} \ket{k_j}$,
with $c_{k,j}$ the amplitude of the $k$-th excited state of mode $j$,
and $\ket{k_j}$ the Fock state with $k$ photons in mode $j$. We
approximate each mode by a harmonic oscillator, changing the 
Hamiltonian from equation~(\ref{eq:MolHamil2Level}) to
\begin{equation}H_{v,0} = \sum_j \omega_j b_j^{\dagger}b_j\, + \omega_a \sigma_z^{(a)}
   + \omega_t a^{\dagger} a \end{equation}

 with $b_j$ being the annihilation operator of the molecule's internal
 vibrational mode $j$ with oscillation frequency $\omega_j$. The
 interaction of the light field $E(t) = E_0 e^{-i \omega_l t}$ with
 the molecule and the motional mode needs to reflect the displacement
 of the motional state with each creation of a vibrational
 phonon. This dynamics is described by the interaction Hamiltonian
\begin{equation} H_I = E(t) \sum_j d_j \left ( e^{-i \Delta_j t} b_j^{\dagger} D(i\eta) + e^{i \Delta t} b_j D(-i\eta) \right )
\label{eq:HamilHarmOsc}
\end{equation}
with $d_j$ being the transition moment, $\Delta_j$ the detuning of the
light field from the molecular vibrational mode $j$, and $E(t)$ the
driving field. It should be noted that the Lamb-Dicke parameter
  $\eta$ is associated with the probe light field frequency and does
  not vary with the vibrational mode.

In general, dynamics governed by the Hamiltonian of equation~(\ref{eq:HamilHarmOsc})
generates a state $\ket{\Psi_{tot}}$ where the vibrational and
motional degrees of freedom are entangled, but mapping the recoil to
the electronic state of the atomic ion does not access the vibrational
state of the molecule. Thus, the information of the vibrational state
can be discarded once the interaction with the ultrafast laser has
ended. Formally, this can be expressed by a partial trace over the
vibrational degree of freedom
$\rho_{motion,tr} = \Tr_{vib} (\ketbra{\Psi_{tot}}) $.  Because the
vibrational and motional degree are entangled, the partial trace
results in a mixed state.

For a single-pulse experiment, the motional state is displaced by the
operator $D(i\eta)$ for each absorbed photon. The traced density
matrix is then derived from the probability to absorb $k$ photons in
mode $j$, expressed by $|\braket{k_j}{\Psi_{mol}}|^2$, from the initial
cat state:
\begin{equation} \rho_{motion,tr}  =  \sum_j \sum_k
  |\braket{k_j}{\Psi_{mol}}|^2 \; D(i \eta k)^\dagger \;
  \ketbra{\Psi_{cat}} D(i \eta k) \; .\end{equation}

As derived in equation~(\ref{eq:det_eff}), the detection probability
is $p_{single} = \sin(2\alpha \eta)^2 $ for a single photon absorption event. The overall detection probability including multiple vibrational
modes is then
\begin{equation}p_{det} = \sin(2\alpha \eta \sum_j \sum_k k_j \,
  |\braket{k_j}{\Psi_{tot}}| )^2 \; . \label{eq:det_prob_mult_modes}\end{equation}

\subsection{Pump-probe absorption spectroscopy: }
\label{sec:pump_probe}

In the following, we extend the described single photon absorption
detection to pump-probe spectroscopy, analyzing the molecule's
vibrational dynamics. This technique relies on the coherent addition
of the individual recoils from each laser pulse.  We introduce a
commonly used model to describe intramolecular vibrational
redistribution (IVR) in the electronic ground state~\cite{Uzer}. We
consider a generic molecule, featuring a strong vibrational infrared
transition from the ground state to the first excited state of the
mode $\nu_b$. The other motional modes $\tilde{\nu}_i$ are dark modes
which couple much weaker to the light field. The anharmonic nature of
the molecular potential leads to a coupling between the dark modes and
the bright mode $\nu_b$.

The quantum state of the entire system is described by the state of
these modes, the combined motional state of the ion crystal and the
atomic electronic state. To improve the clarity of the notation, we
separate the state of the bright mode
$\ket {\nu_b}=\sum_k c_k \ket{k_b}$ from the dark modes
$\ket{\tilde{\nu}}$. The initial state of the system has all
vibrational modes in the ground state and the combined motional and
atomic electronic degrees of freedom are in the state
$\ket{\Psi_{cat}}$. The combined state is then expressed as:
\begin{equation}\ket{\Psi_{tot}} = \ket{0_b}\otimes \ket{\tilde{0}} \otimes \ket{\Psi_{cat}}
\end{equation}
with the ground state of all dark modes being $\ket{\tilde{0}} = \otimes_{j \neq b}\ket{0_j}$.

For pump-probe spectroscopy, two infrared pulses are sent onto the
molecule whose recoil can be described by identical displacement
operators. The pump pulse prepares the bright mode in a
well-defined state.
For simplicity, we assume that the first (pump) pulse of the
experiment prepares the molecule perfectly in the first excited state
of the bright mode $\ket{\nu_b}= \ket{1_b}$. Thus, the bright mode can
be modeled as a two-level system, altering the molecule-light
interaction Hamiltonian from equation~(\ref{eq:HamilHarmOsc}) to:
\begin{equation} H_I = E(t)  d_b \left [ e^{-i \Delta_b t} \sigma_+^{(b)} D(i\eta) + e^{i \Delta_b t} \sigma_-^{(b)} D(-i\eta) \right ]
\end{equation}
where the Pauli operators with index $(b)$ acts on the bright mode.

A well-calibrated pump pulse induces thus a single vibrational quantum
in $\nu_b$, corresponding to a single photon absorption event,
displacing the motional state of the ion-crystal:
\begin{equation}\ket{\Psi_{pump}} = \ket{1_b}\ket{\tilde{0}} \otimes D(i\eta)\ket{\Psi_{cat}} \; .
\end{equation}

The system is then subject to a free evolution with time $t$. During
this time, the vibrational excitation is redistributed over the
dark modes $\tilde{\nu}_m$, yielding the state
\begin{equation}
  \ket{\Psi_{evol}} = c_0(t) \ket{1_{b}}\otimes \ket{\tilde{0}}  \otimes D(i\eta)\ket{\Psi_{cat}} + \sum_m c_m(t) \ket{0_{b}}\otimes \ket{\tilde{\nu}_{m}} \otimes D(i\eta)\ket{\Psi_{cat}}
\end{equation}
where the states $\ket{\tilde{\nu}_m}$ form a basis of the dark modes.

Then, the second (probe) pulse is applied onto the molecule, aiming to
undo the pump pulse, de-exciting the bright mode $\nu_b$. For the part
of the state where $\nu_b$ is still in the excited state $\ket{1_b}$,
the pulse de-excites the vibrational mode into the ground state. This
de-excitation comes along with the stimulated emission of a photon
which causes a displacement with opposite sign relative to the pump
pulse $D(-i \eta)$. For the part of the state where the population
from the bright mode has been transferred to a dark mode (and thus the
$\nu_b$ mode is in its ground state $\ket{0_b}$), the probe pulse 
causes another absorption event, exciting the $\nu_b$ mode, and causing
a displacement that is in phase with the displacement from the pump
pulse $D(i \eta)$.

The state of the system after the probe pulse at time $t=\tau$ is then given by

\begin{equation}\label{eq:pump_probe_theory}
\fl  \ket{\Psi_{final}}  = c_0(\tau) \ket{0_b} \otimes \ket{\tilde{0}} \otimes \ket{\Psi_{cat}} + 
   \sum_m \tilde{c}_m(\tau) \ket{1_b}\otimes \ket{\tilde{\nu}_m} \otimes
  D(2 i\eta) \ket{\Psi_{cat}} \, .
\end{equation}

The exciting and de-exciting momentum kicks cancel each other if the
vibrational excitation remains in the $\nu_b$ mode, expressed by the
amplitude $c_0(\tau)$. For the other parts of the state, the momentum
kicks add coherently, resulting in a displacement of
$D(2 i\eta)$.

The displacement after both pulses is then measured with the recoil
detection method outlined in section~\ref{sec:absorption_spec}.
The cancellation of the momentum kicks relies on
the fact that the displacement operations caused by the pump pulse and
the probe pulse have opposite phases. This assumption is only valid if
the harmonic oscillator of the motional state does not evolve in the
time between the pump and the probe pulses. In typical molecular
systems, the vibrational dynamics occurs at the picosecond timescale
which is short compared to the oscillation period of the ion crystal
in the microsecond range. Thus, it can be assumed that the harmonic
oscillator does not evolve between the pump and the probe pulse.

\subsection{Relevant molecular properties and experimental requirements}
\label{sec:molecule_prop}\label{sec:exper-requ}

The presented quantum logic methods are independent of the molecular
properties but several practical considerations need to be taken
  into account:
\begin{itemize}
\item It should be possible to ionize the molecule without
  fragmentation and the cation should be stable. 
\item The molecule needs to be prepared in its vibrational
  ground state.
\item The molecules should be stable under ion-trapping
  conditions. Electric fields used to manipulate and cool the atomic
  logic ion should not dissociate or affect the molecular structure.
\item The ratio between the mass of the molecule and the mass of the
  atomic logic ion should be within a factor of two. This allows
  for efficient sympathetic cooling and quantum logic readout.
\item The molecule should feature vibrational modes with a
  strong infrared transition that is accessible with available laser
  systems. Accessible frequencies lie in the range from 4000\wavenum
  to 1500\wavenum. 
\end{itemize}

In the following, we analyze the recoil detection protocol for the
cations of Ammonia (NH$_3^+$), Acetylene (C$_2$H$_2^+$),
Cyanoacetylene (C$_3$HN${^+}$), Aniline (C$_6$H$_5$NH$_2{^+}$), and
Phenylalanine (C$_9$H$_{11}$NO$_2^+$). These ionic molecules have been
spectroscopically investigated in their gas
phase~\cite{Rizzo2009,Hojbjerre2008,Desrier2017,Schlemmer2005}. The
set of molecules spans a mass range from 17 to 165 Dalton and thus
multiple logic ion species need to be considered, in particular \Ca,
\Sr, and \Ba. These atomic ions are well suited for co-trapping
molecular ions, as their internal states are manipulated by a
quadrupole transition at wavelengths of 729nm, 674nm, and 1.7$\mu$m
respectively. In contrast, lighter atomic ions such as $^9$Be$^+$ and
$^{20}$Mg$^+$ require UV-Raman lasers to manipulate their electronic
state, which might dissociate the
molecule~\cite{Wineland1995Experimental}.

In the following, an overview over a suitable experimental setup is
given.  It is expected that motional decoherence is the main source of
errors for the recoil detection~\cite{Hempel2013}.  A macroscopic
linear quadrupole ion trap is the natural choice due to its large
ion-surface distance that allows for low motional heating rates and
thus excellent motional coherence. Heating rates below a single quantum/s
have been observed in macroscopic traps with electrode-ion distance at
the millimeter scale~\cite{Brownnutt2014}. It is expected that the
motional coherence can be increased by more than one order of
magnitude when cooling the trap to 70K, as it is known that the
surface processes that lead to excessive motional heating are
suppressed at low temperatures~\cite{Brownnutt2014}. Thus, a
macroscopic linear trap at a temperature of 70K should achieve a
heating rate of significantly less than 1 quanta/s.

A simple way to co-trap the particles is to ionize the molecule inside
the trapping volume, where the logic ion is already trapped and
cooled. This can be achieved, by injecting the molecules in the gas
phase into the vacuum vessel and ionizing them with a suitable
method. The external motion of the molecule is immediately
cooled by the logic ion after the ionization which leads to the
formation of a Coulomb crystal which can be detected by a change in
position of the logic-ion. Robust molecules can be readily ionized by
electron bombardment, whereas more complex and fragile molecules require
multi-photon ionization techniques~\cite{VieiraMendes2012}.

\begin{figure}[ht]
\begin{center}
  \includegraphics[width=.8 \textwidth]{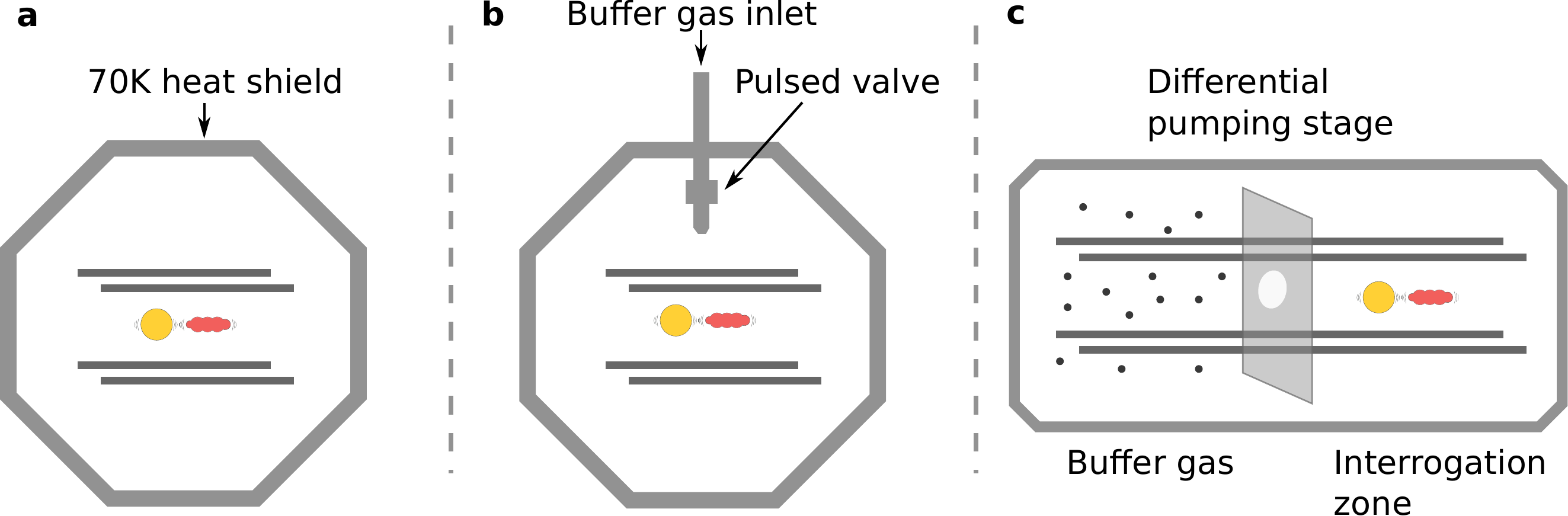}
  \caption{\label{fig:exp_setup} Sketch of three possible trapping
    configurations. a) Cryogenic trapping setup relying on black body
    radiation to internally cool the molecule. b) Simple Paul trap
    setup with pulsed Buffer gas cooling. c) Multi-zone trapping
    apparatus with differential pumping to ensure the required high
    vacuum in the interrogation zone.}
\end{center}
\end{figure}

The vibrational state of the molecule needs to be cooled close to its
ground state after the ionization process. Furthermore, energy that is
introduced by the interrogation pulses needs to be removed.  A single
trapped molecule in a high vacuum environment undergoes effectively no
collisions that could remove this excess energy. Thus, black-body
radiation is the only mechanism that is available. For strong
infrared vibrational transitions the radiative lifetime is on the
millisecond to second timescale. Thus, the repetition rate of any
experiment relying on black-body radiation is limited by this
radiative lifetime. A possible setup for a cryogenic ion trap relying
on black body radiation is shown in figure~\ref{fig:exp_setup}a.

Alternatively, the internal states of the molecule can be cooled by
collisions with a cold neutral buffer gas~\cite{Hansen2014}. The
cooling process can be described by Langevin collisions of the hot
molecule with the cold buffer gas~\cite{Hutzler2012}. The required
cooling time depends on the initial temperature of the molecule, the
mass-ratio between the molecule and the buffer gas, and the pressure
of the buffer gas~\cite{Hutzler2012}. Typical cooling times for buffer
gas cooling in Paul traps are in the order of a few
milliseconds~\cite{Hutzler2012,Dzhonson2006}.

However, the recoil detection techniques needs to be performed under
high vacuum conditions, as continuous buffer gas cooling would disturb
the detection procedure. Following techniques for selective Buffer gas
cooling are available: (i) A pulsed buffer gas source as demonstrated
in reference~\cite{Vazquez2018} and sketched in
figure~\ref{fig:exp_setup}b. (ii) A multi-zone ion trap with multiple
differential vacuum levels, similar to the proposed ion conveyor belt
of reference~\cite{Kahra2012}, sketched in
figure~\ref{fig:exp_setup}c. In this case, the ion crystal needs to be
transferred between the two different zones between each experimental
cycle. Such transfer operations are routinely performed in segmented
ion traps designed to enable scalable quantum information
processing~\cite{Kaufmann2017}.

Control over the molecule's vibrational state is provided by an
infrared ultrafast femtosecond laser system. The laser system needs a
sufficiently high light intensity to perform the required operations
with a single laser pulse on the vibrational mode. A pulse energy of
600nJ, yielding a peak laser power of $3 \cdot 10^6$W focused to a
spot size of 10$\mu$m, results in a Rabi frequency of approximately
$2 \pi \, 4 \cdot 10^{13}$Hz for a strong infrared transition. This
Rabi frequency should be sufficient to transfer all population into
the first excited state of a strong vibrational transition with a
single laser pulse.

Such a laser system is commercially available using a high power
fiber-based master oscillator and an optical parametric amplifier.
This produces laser pulses at wavelengths ranging from 3\mum to
10\mum with a duration of 200fs and a single pulse energy ranging from
800nJ to 200nJ. The repetition rate of the system is 1MHz which can be
synchronized with the frequency of the motional mode of the trapped
ions in order to maximize the acquired geometrical phase in the
absorption detection protocol.
Single pulses can be selected using an acousto-optical
modulator.

\subsection{Analysis of recoil detection}
\label{sec:expect-exper-perf}

In the following, we analyze the expected performance of the recoil
detection with realistic experimental parameters. First, we review
the generation procedure of the entangled cat state. The electronic
and motional degrees of freedom of the atom can be described by the
Hamiltonian that has been introduced for a two-level molecular system
in equation~(\ref{eq:LightAtomHamil}). There, the vibrational frequency
$\omega_v$ of the molecule needs to be replaced with the atomic
electronic transition frequency $\omega_a$, and the value of the
Lamb-Dicke parameter needs to be adapted accordingly.

The interaction Hamiltonian from equation~(\ref{eq:LightAtomHamil})
can be simplified by expressing it in a rotating frame oscillating
with the two-level system's transition frequency, using rotating wave
approximation, and performing the Lamb-Dicke approximation
$\exp(i \eta (a+a^\dagger)) \approx 1 + i \eta (a+a^\dagger)$. The
simplified interaction picture Hamiltonian is then
\begin{equation}H_{int} = \frac{\hbar \Omega}{ 2} (e^{-i\Delta t} \sigma_+^{(a)} (1+i\eta (a e^{-i \omega_t t} + a^\dagger e^{i \omega_t t}) ) + \mathrm{H. \, c.}) 
\label{eq:LDHamil}
\end{equation}
with the detuning $\Delta = \omega_l - \omega_a$.  The Lamb-Dicke
approximation is valid for $\eta (2\bar{n}+1) \ll 1$ with $\bar{n}$
being the mean phonon number of the motional mode.

For the generation of the desired cat state, a bichromatic field is
applied to the atom where the two tones are detuned by the motional
trap frequency, i.e. $\Delta_\pm = \pm w_t$. When neglecting terms
oscillating with frequency $\omega_t$ or higher, the interaction with
the two light fields can be expressed by
\begin{equation}H_\pm = \frac{i \hbar \eta \Omega}{ 2} (a \sigma_\pm^{(a)} - a^\dagger \sigma_\mp^{(a)}) \; .
\end{equation}
Combining the two light fields and using the identity
$\sigma^{(a)}_x = \sigma^{(a)}_+ + \sigma^{(a)}_-$ leads to the interaction Hamiltonian
\begin{equation}H_{bic} = \frac{\hbar \eta \Omega}{2} \sigma_x^{(a)} (a+a^\dagger)  \, .
\end{equation}
This Hamiltonian displaces the state of the
harmonic oscillator in a direction in phase space that depends on the
expectation value of the electronic $\sigma_x^{(a)}$ operator. The ground state of the
atomic electronic state $\ket{g}_a$ can be expressed as a
superposition of the two eigenstates of the $\sigma_x^{(a)}$ operator
$\ket{g}_a=1/\sqrt(2)( \ket{+}_x + \ket{-}_x) $. Applying the
bichromatic light field to the ground state of the atomic electronic
state and the motional state  $\ket{g}_a \otimes \ket{0}$ for a duration $t$
generates thus the entangled state
\begin{equation} \ket{\Psi_{cat}} = \frac{1}{\sqrt{2}} \biggl{(}\ket{+}_x \otimes \ket{\alpha} + \ket{-}_x \otimes \ket{-\alpha}\biggr{)}
\end{equation}%
with the displacement $\alpha=i \eta \Omega t / 2$.

In order to reach appreciable single-photon detection efficiency (see
equation~(\ref{eq:det_eff}), cat states with displacement of
$\alpha > 10$ are required. In this regime, the Lamb-Dicke
approximation from equation~(\ref{eq:LDHamil}) is no longer
valid. Mathematically, the Lamb-Dicke approximation is a first order
Taylor expansion of $\exp(i\eta (a+a^\dagger))$. A similar Hamiltonian
can be retrieved by keeping all orders of $\eta$ and applying the rotating
wave approximation, neglecting fast oscillating
terms~\cite{leibfried_rmp_2003}. In the resulting Hamiltonian, the
lowering operators $a$ are replaced by the operators $\mathcal{A}$ which,
applied to a Fock state with phonon number $n$, yield
\begin{eqnarray}\mathcal{A} \ket{n} = F(n) \ket{n-1} \\
    \mathrm{with} \, F(n) = e^{- \eta^2/2}  \sqrt{1/(n+1)} L_n^{1}(\eta^2)   %
\end{eqnarray}
with $L_n^{1}(x)$ being the generalized Laguerre polynomial. This
leads to a coupling strength that is reduced for large
phonon numbers as compared to the Lamb-Dicke approximation.  With this
operator, the modified interaction picture Hamiltonian in all orders
Lamb-Dicke can be expressed as:
\begin{equation}H_{int} = \frac{\hbar \Omega}{ 2} (e^{-i\Delta t} \sigma^+ (1+i\eta (\mathcal{A} e^{-i \omega_t t} + \mathcal{A}^\dagger e^{i \omega_t t}) ) + \mathrm{H. \, c.}) \; . 
\label{eq:AllHamil}
\end{equation}
With this Hamiltonian, the cat state preparation procedure can be
recovered by applying the same bichromatic light field. For cat states
with large displacements, the generation time is increased and there
is a maximum possible displacement of the cat state, as the coupling
to the sideband vanishes for $L_n^{1}(\eta^2)=0$, halting the
generation dynamics. Higher maximum displacements can be reached by
reducing the Lamb-Dicke parameter, which can be achieved by increasing
the overlap angle $\theta$ between the motional mode and the
wavevector of the light field (see equation~\ref{eq:LDDef}). However,
reducing the Lamb-Dicke parameter slows the overall state generation,
making the process more vulnerable to decoherence.

In practice, the decoherence of the motional mode of the ion crystal
is likely limited by motional heating of the ion
crystal~\cite{Brownnutt2014}.  The sensitivity of the detection
protocol to motional heating has been investigated in
reference~\cite{Hempel2013}, yielding an error model where the
motional decoherence is modeled by random phase fluctuations
$\langle \phi_h^2 \rangle$. These phase fluctuations are estimated
by
\begin{equation}\langle \phi_h^2 \rangle = 8 R_h \alpha^2 \frac{2 \tau}{3}
\end{equation}
for a cat state with displacement $\alpha$, generation duration
$\tau$, and heating rate $R_h$. The coherence of the Ramsey-like
sequence to detect the geometric phase from the photon absorption
event is reduced from the ideal value by $C = \exp(-\langle \phi_h^2 \rangle/2)$.

\begin{table}
  \begin{centering}
  \begin{tabular}{cccc}
    \hline \hline
    Mode & Experiment \cite{Smith-Gicklhorn2001} & Experiment~\cite{Desrier2017} & Theory \\ \hline
    $\nu_1$ & 3196.5 (160) & 3123 & 3259 (213)\\
    $\nu_2$ & 2175.8 (31) & 2177 & 2206 (2)\\
    $\nu_3$ & 1852.8 (372)& 1855 & 1890 (334)\\
    $\nu_4$ &  - & 829 & 911 (7)\\ \hline
  \end{tabular}
  \caption{\label{tab:vibration}Experimental and theoretical vibrational frequencies of the
    vibrational stretching modes of the Cyanoacetylene cation. Theoretical values have been calculated
    with restricted open shell DFT calculations at the b3lyp level and
    a scale factor of 0.97. All values are given in \wavenum. Values
    in brackets are IR-intensities in km/mol.  }
  \end{centering}
\end{table}

\begin{figure}[ht]
\begin{center}
  \includegraphics[width=.45\textwidth]{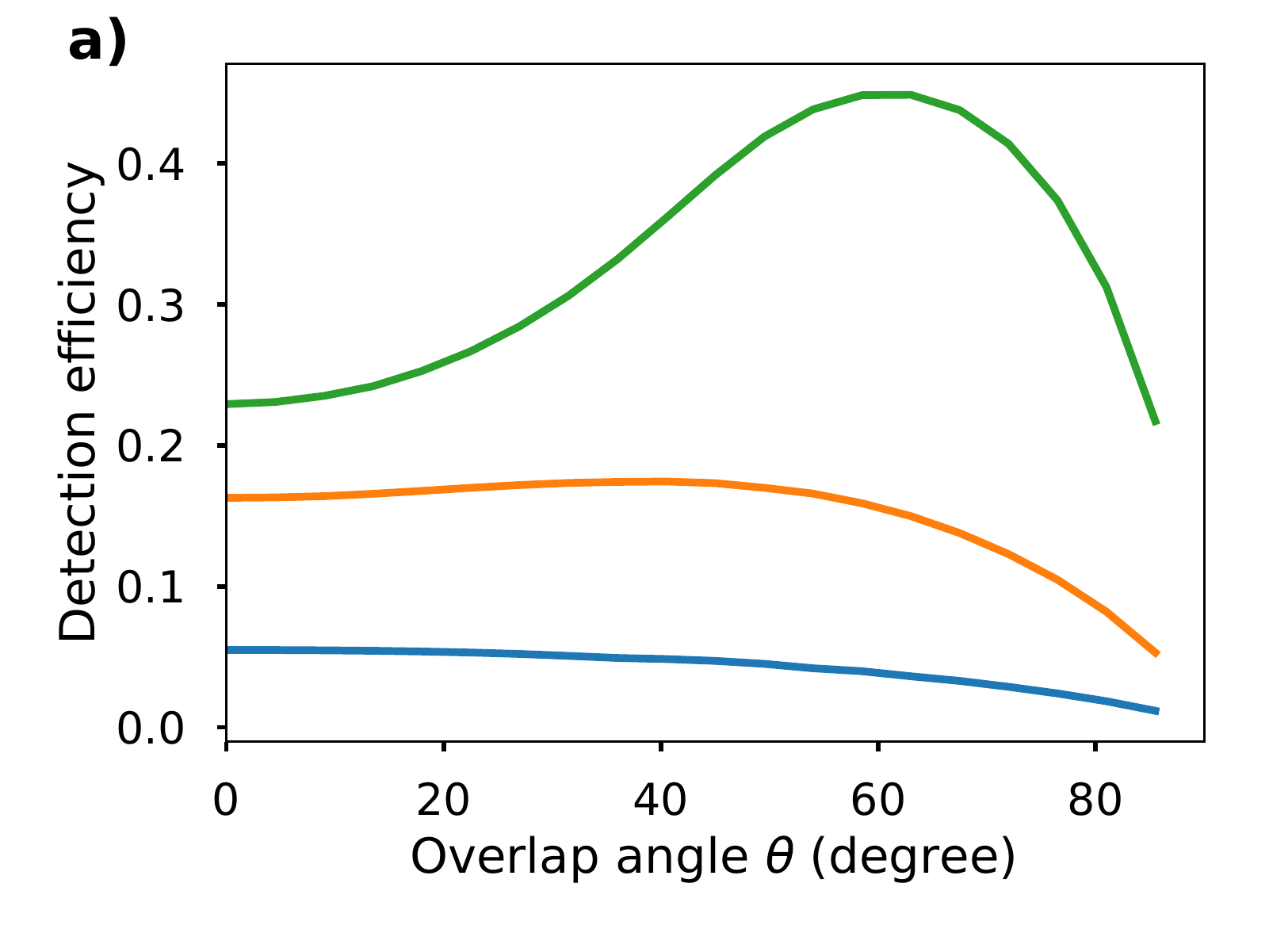}
    \includegraphics[width=.45\textwidth]{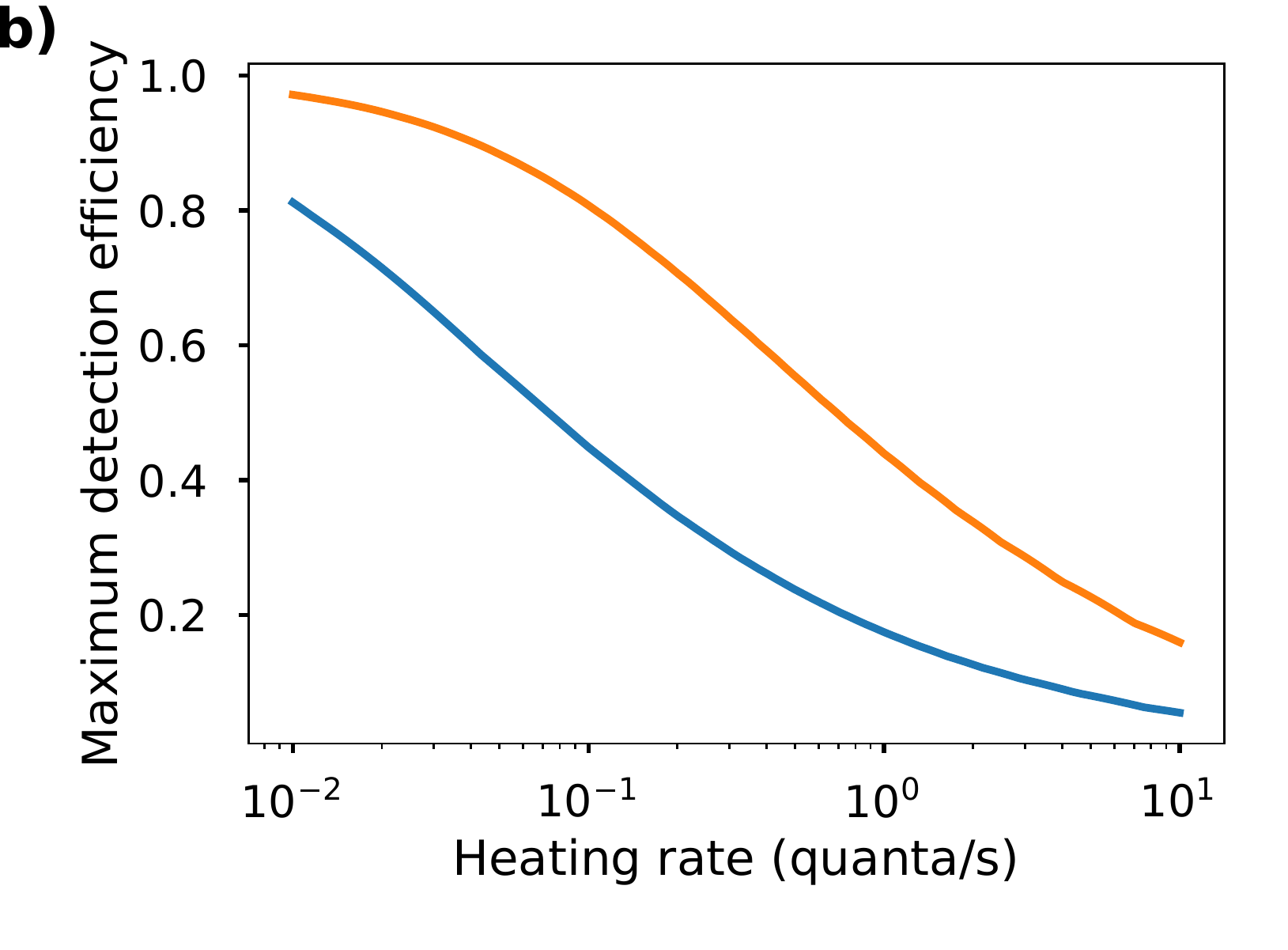}
    \caption{\label{fig:single_shot_prob} (a) Probability to detect a
      single-photon absorption event on the $\nu_3$ mode of
      Cyanoacetylene as a function of the overlap angle between the
      laser wavevector and the motional mode for a heating rate of
      0.1,~1,~10 quanta/s, shown in green, orange, and blue
      respectively.  (b) Maximum single-photon detection efficiency as
      a function of heating rate, optimized over the overlap angle for
      the $\nu_1$ and $\nu_3$ mode shown in orange and blue
      respectively.}
\end{center}
\end{figure}

In order to reach the optimum detection efficiency for a given
motional heating rate, the Lamb-Dicke parameter needs to be adapted by
choosing a geometry with suitable angle $\theta$ between the wave
vector and the motional mode. We demonstrate this optimization for the
Cyanoacetylene cation and a \Ca logic ion. The vibrational transition
frequencies and their respective strengths for Cyanoacetylene are
shown in table~\ref{tab:vibration}. In particular, the vibrational
modes $\nu_1$ and $\nu_3$, with transition frequencies of 3259\wavenum
and 1890\wavenum, show a high dipole moment and are thus of
interest. The recoil detection procedure is analyzed
numerically~\cite{Johansson2013}, where the Rabi frequency on the
atomic transition is limited by the available laser power to
$\Omega = 2\pi \, 300\,$kHz~\cite{Schindler2013}.  The probability to
detect a single photon absorption event on the $\nu_3$ mode as a
function of the overlap angle $\theta$ and multiple motional heating
rates is shown in figure~\ref{fig:single_shot_prob}a. It can be seen
that the heating rate determines the optimal angle. This is expected,
as reducing the Lamb-Dicke parameter (larger angle) gives access to
larger displacements but leads to a slower state generation that is
more affected by the decoherence. The maximum achievable detection
efficiency for a heating rate of 0.1 quanta/s is about 49\%. The
maximum achievable detection efficiency as a function of the heating
rate, is shown in figure~\ref{fig:single_shot_prob}b for the two
strong stretching modes $\nu_1, \nu_3$ of Cyanoacetylene. It can be
seen that the detection efficiency of the $\nu_1$ mode is larger due
to the higher mode frequency resulting in a larger recoil per absorbed
photon.

\begin{figure}[ht]
\begin{center}
\includegraphics[width=.55 \textwidth]{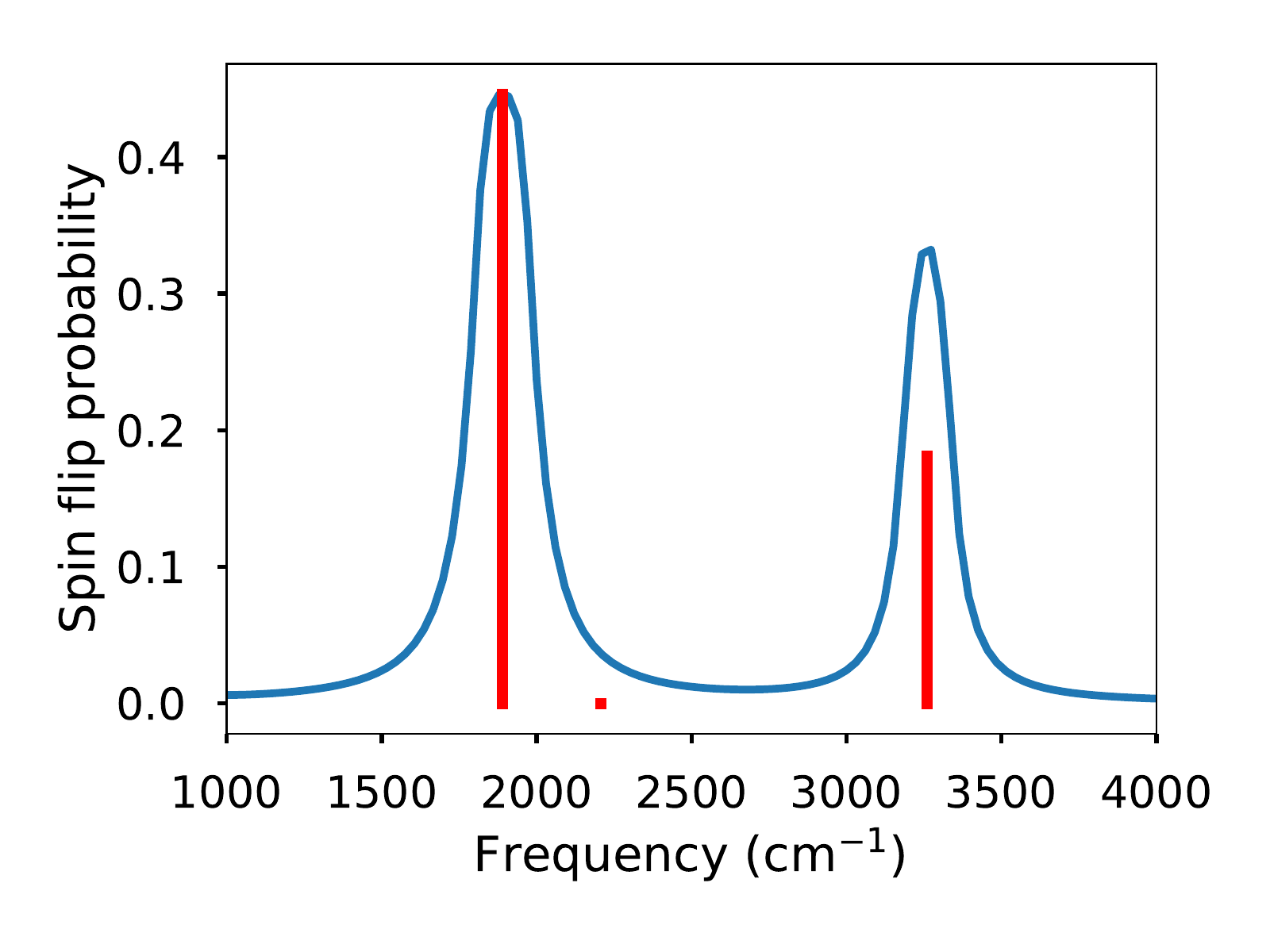}
\caption{\label{fig:abs_many} Spin-flip probability of the atomic
  ion when performing spectroscopy on a co-trapped Cyanoacetylene
  molecule. The red bars indicate the position and relative oscillator
  strengths of the individual vibrational modes. }
\end{center}
\end{figure}

With this analysis, one can infer the expected spectrum when applying
a singe 200fs laser pulse onto a single Cyanoacetylene cation from
equation~(\ref{eq:det_prob_mult_modes}).  Figure~\ref{fig:abs_many}
shows the expected atom signal for a heating rate of 0.1~quanta/s,
where the red bars indicate the relative oscillator strength of the
individual modes. It can be seen that the expected spin-flip
probabilities do not follow the relative oscillator strengths, because
the Lamb-Dicke parameter and thus also the detection efficiency is
increased for larger mode frequencies.

\begin{table}
  \begin{centering}
  \begin{tabular}{ccccccccc}
    \hline \hline
    Molecule & Mass & Vib. & Frequency &Logic       & \multicolumn{3}{c}{Detection efficiency}  \\ 
             & (Dalton)  & mode  & (\wavenum)  &Ion & 10&1&0.1\\ \hline
    NH$_3^+$& 17 & $\nu_1$ NH str & 3498 & \Ca & 30\%  &67\%& 94\%\\
    C$_2$H$_2^+$ & 26 & $\nu_2$ CH str & 3363 & \Ca & 25\% &59\% & 91\%\\
    C$_3$HN${^+}$ / $\nu_3$ &  51 & $\nu_3$ & 1890 & \Ca & 6\% &20\%&50\% \\
    C$_3$HN${^+}$ / $\nu_1$& 51 & $\nu_1$ CH str & 3259 & \Ca & 16\%& 44\% &81\% \\
    C$_6$H$_5$NH$_2^+$& 93 & $\nu_2$ NH str & 3398 & \Sr & 9\% & 30\% & 66\%\\
    C$_9$H$_{11}$NO$_2^+$ & 165 & $\nu_3$ NH str & 3382 & \Ba & 3\% &11\%&40\% \\
    \hline
  \end{tabular}
  \caption{\label{tab:AllMolecules}Expected detection efficiencies for
    heating rates 10,~1,~0.1~quanta/s for the molecules Ammonia
    (NH$_3$), Acetylene (C$_2$H$_2$), Cyanoacetylene (C$_3$HN${^+}$),
    Aniline (C$_6$H$_5$NH$_2{^+}$), and Phenylalanine
    (C$_9$H$_{11}$NO$_2^+$). The values are estimated for a motional
    mode frequency of 500kHz and a maximum Rabi frequency of
    300kHz. Vibrational frequencies have been calculated with
    restricted open shell DFT calculations at the b3lyp level and a
    scale factor of 0.97.}
  \end{centering}
\end{table}

The detection efficiency has been analyzed for multiple molecular
species, requiring only the mass and the frequency of the relevant
vibrational modes for each molecule. Here, mostly CH- and
NH-stretching modes with a frequency of around 3500\wavenum have been
considered. Table~\ref{tab:AllMolecules} shows the vibrational
transition frequencies of the investigated molecular ions including
suitable atomic logic ions that are chosen depending on the molecule's
mass. The expected detection efficiency for each transition is
estimated for heating rates of 10,~1,~0.1~quanta/s in
table~\ref{tab:AllMolecules}. It can be seen, that light molecules, in
a system with low-heating rates, allow near perfect detection of a
single-photon event. Even for comparatively heavy molecules, such as
Phenylalanine with a mass of 165 Dalton, a single photon absorption
detection efficiency of 40\% seems achievable.

\subsection{Analysis of pump-probe spectroscopy}
\label{sec:expect-exper-pump-probe}

Finally, we investigate the expected outcomes of pump-probe
experiments, yielding information about intramolecular dynamics.
We assume a simple irreversible IVR model where the population from
the bright state decays into the dark states with rate
$\tau_1$~\cite{Uzer}. The expected momentum transfer is shown in
equation~(\ref{eq:pump_probe_theory}), where the time-dependent
population of the bright state needs to be set to
$c_0(\tau)^2 = \exp(-\tau/\tau_1)$. The population that has leaked to
a dark state, expressed by $c_b^2 = 1-c_0(\tau)^2$, adds a momentum
kick corresponding to two photon absorption events $D(2i\eta)$. This
increase in displacement magnitude leads to higher detection
efficiencies than the single pulse spectroscopy discussed above. The
expected detection efficiencies $p_{2 photon}$ for the investigated
molecules are shown in table~\ref{tab:AllMoleculesPump}. The expected
spin-flip probability for the pump-probe experiment  $p_{pump/probe}$
as a function of the laser pulse delay $\tau$ can then be written as:
\begin{equation}
p_{pump/probe} = (1-\e^{-\tau/\tau_1)} \, p_{2 photon}
\end{equation}
\begin{table}
  \begin{centering}
  \begin{tabular}{cccc}
    \hline \hline
    Molecule & \multicolumn{3}{c}{Detection efficiency} \\
              & 10 & 1 & 0.1  \\ 
    \hline
    NH$_3^+$& 70\% & 95\% & 99\% \\
    C$_2$H$_2^+$ & 64\% & 93\% & 99\% \\
    C$_3$HN${^+}$ / $\nu_3$ & 22\% & 57\% & 89\% \\ 
    C$_3$HN${^+}$ / $\nu_1$ & 46\% & 81\% & 98\% \\
    C$_6$H$_5$NH$_2^+$& 29\% & 72\% & 95\% \\
    C$_9$H$_{11}$NO$_2^+$ & 10\% & 36\% & 79\% \\
    \hline
  \end{tabular}
  \caption{\label{tab:AllMoleculesPump}Expected detection efficiencies
    for the pump-probe spectroscopy given heating rates
    10,~1,~0.1~quanta/s for the molecules Ammonia (NH$_3$), Acetylene
    (C$_2$H$_2$), Cyanoacetylene (C$_3$HN${^+}$), Aniline
    (C$_6$H$_5$NH$_2{^+}$), and Phenylalanine
    (C$_9$H$_{11}$NO$_2^+$). The respective transition frequencies are
    shown in table~\ref{tab:AllMolecules}. The values are estimated
    for a motional mode frequency of 500kHz and a maximum Rabi
    frequency of 300kHz. }
  \end{centering}
\end{table}

The presented analysis of the pump-probe spectroscopy assumes that a
two-level system of the two lowest energy motional modes can be
isolated. In the limit of small anharmonic shifts, the vibrational
modes need to be treated as a harmonic oscillators and thus also
higher excited states are populated by a single laser pulse. It has
been shown in section~\ref{sec:absorption_spec} that the detection
procedure is not influenced in the limit of small anharmonic shifts.

However, the IVR dynamics of higher excited vibrational modes might
differ substantially from the dynamics of the first excited
state. This is due to the fact that the IVR timescale is determined by
the density of the dark modes which usually differs from the first to
second excited state~\cite{Uzer}. Thus, a clean IVR signal requires
one to populate only the first excited state of the bright
mode. Optimal control methods can provide a temporal amplitude profile
for the laser pulse that isolates a two-level system even for small
anharmonic shifts~\cite{Berrios2012}.

\section{Discussion and outlook}

We have introduced non-destructive methods to perform ultrafast
spectroscopic techniques on single molecular ions using quantum logic
readout with a co-trapped atomic ion. We anticipate that a single
photon absorption event can be detected with probabilities of larger than 90\% in
lighter molecules with mass less than 50~Dalton. The detection
efficiency can reach up to 40\% in heavier molecules with mass around
150~Dalton. We have developed a pump-probe technique based on this
absorption detection that can give insight into intramolecular
dynamics of single molecular ions.

Within this manuscript we have deliberately chosen a simple model of
the molecule to emphasize the basic working principle. We briefly
discuss the expected impact of a more complex molecular model
including additional degrees of freedom:
\begin{itemize}
\item Rotational degree of freedom: The rotational degree of freedom
  of the molecules has been neglected in the light-molecule
  interaction. This approximation is justified because the laser
  pulses are usually in the femtosecond timescale and rotational
  dynamics of polyatomic molecules occurs in the picosecond to
  nanosecond domain. Thus, rotational effects will play a role when
  investigating the time dynamics of the vibrational modes but will
  not affect the spectroscopic technique itself. The pump-probe
  experiments are expected to show rotational dephasing and rephasing
  dynamics as have been observed in ultrafast experiments with neutral
  molecules~\cite{Lemeshko2013}.
\item Spin-orbit coupling and Zeeman substructure: For the
  investigated Cyanoacetylene molecule, the spin-orbit coupling is
  much smaller than the spectral width of the exciting femtosecond
  laser pulses~\cite{Desrier2017}. This substructure cannot be
  resolved and can thus be neglected for the light-molecule
  interaction with ultrafast light pulses.
\end{itemize}

The presented methods are stepping stones towards more complex
spectroscopic schemes. Existing spectroscopic methods can be
transferred using the total recoil of all applied laser pulses as the
measured quantity.

The amount of extracted information from the molecule can be increased
by exploiting multiple modes of the motion of the ion crystal. This
will give access to the correlations between the pathway and the final
state of intramolecular processes on a single-molecule and single-shot
level. For example, the first mode can be used to measure the final state of
the dynamics and the second mode can be used to distinguish distinct
pathways of intramolecular dynamics by introducing state-dependent
momentum kicks, similar to ultrafast gate operations in atomic
ions~\cite{Garcia-Ripoll2003}.

\section*{Acknolwedments}
The author acknowledges support from the Austrian Research Promotion
Agency (FFG) contract 872766 and likes to thank the anonymous referees
for their valuable comments.

\clearpage
\bibliography{start}
\end{document}